\documentclass[12pt]{article}
\usepackage{amsmath,amssymb} 

\newcommand{\ket }{\rangle }
\newcommand{\bra }{\langle }

\begin{document}
\begin{flushright} 
NUP-A-98-15 \\ Sept. 1998
\end{flushright}
~\\ ~\\ 
\begin{center}
\Large{\bf{Physics Involved in Neutrino Oscillation}}
\end{center}
~\\ ~\\ 
\begin{center}
Yoshihiro TAKEUCHI, Yuichi TAZAKI, S. Y. TSAI\footnote{E-mail address: tsai@phys.cst.nihon-u.ac.jp} and Takashi YAMAZAKI
\end{center}
\begin{center}
{\it Atomic Energy Research Institute and Department of Physics \\ 
College of Science and Technology, Nihon University \\ 
Kanda-Surugadai, Chiyoda-ku, Tokyo 101, Japan }
\end{center}
~\\ ~\\ ~\\ 

%%% Abstract %%%

\begin{abstract}
  Neutrino oscillation is discussed with emphases placed more on its conceptual aspects. After reviewing the two conventional formulations, referred to here as the same-energy prescription and the same-momentum prescription, wave packet treatments are developed for each of these two prescriptions. Both wave packets localized in space and those in time are considered, and, by invoking relativistic kinematics as well, the necessary conditions for oscillation to occur are derived, which appear to have a form more well-defined and quantitative than what have been noted before. Some phenomenological implications suggested by the wave packet treatments are briefly mentioned. Finally, as a possible third prescription, the same-velocity prescription is given.
\end{abstract}

%%%%%%% S 1: INTORO %%%%%%%%%%
\newpage 

\section{Introduction}

  Neutrino oscillation$^1$ is one of the hot topics in particle physics, non-accelerator as well as accelerator high energy physics and astrophysics. Evidence in favor of oscillation in atmospheric neutrinos was reported by the Super-Kamiokande Collaboration in the International Conference Neutrino 98 held recently in Takayama, Japan,$^2$ which has attracted much attention even beyond the academic communities.$^3$ A number of papers have followed, most of which are more or less phenomenology and/or model-oriented.$^4$ In contrast, the emphases of the present note will be placed more in conceptual aspects of neutrino oscillation.

  As well known, the main physics involved in neutrino oscillation is quantum mechanics and this aspect of neutrino oscillation has once been discussed in some detail in a paper by Kayser,$^5$ from which we have learned much. We are going to elaborate and extend his arguments and treatments, in paticular, to appreciate explicitly relevance of wave packet treatments and relastivistic kinematics, and to reveal thereby physics involved in a more instructive and comprehensible way.

  This note is arranged as follows. In Sec.II which reviews conventional formalism of neutrino oscillation, the two prescriptions, to be referred to respectively as same-energy prescription and the same-momentum prescription are introduced. In Sec.III, wave-packet treatments are developed and conditions for neutrino oscillation to occur are derived. In Sec.IV, some phenomenological implications are briefly mentioned. In Sec.V, as a possible third prescription, the same-velocity prescription is given. Sec.VI devotes to concluding remarks.

%%%%%%% Sect 2.  %%%%%%

\section{Conventional formalism}

  Let $|\nu_f \ket $ ($f = e$, $\mu$, $\tau$, etc.) be flavor eigenstates, i.e. the neutrinos associated with electron, muon, $\tau$ lepton, etc., and suppose that they are mutually-orthogonal superpositions of the mass eigenstates $|\nu_i \ket $ having mass $m_i$ ($i = 1, 2, 3,$ ... ): 
%%%(1)
\begin{equation}
|\nu_f\ket =\sum_i U_{fi}|\nu_i\ket ~.
\end{equation}
We shall suppose that there are $N$ flavor eigenstates and $N$ mass eigenstates with their masses arranged as 
%%%(2)
\begin{equation}
m_1 < m_2 < \dots < m_N ~,
\end{equation}
and, assuming time reversal invariance, take the $N\times N$ mixing matrix $U=(U_{fi})$ to be a real orthogonal matrix.

  Suppose a neutrino is born with definite flavor $f$ at $x=0$ and $t=0$, then its state vector at $x$ and $t$ may be written as
%%%(3)
\begin{equation}
|\nu_f(x,t)\ket =\sum_i U_{fi}e^{i(p_ix-E_it)}|\nu_i\ket ~,
\end{equation}
where
%%%()
\[
E_i=\sqrt{p_i^2+m_i^2}~.
\]
The probability to find neutrino with flavor $f'$ at $x$ and $t$ is calculated as
%%%(4)
\begin{eqnarray}
P_{f\to f'}(x,t)&=&|\bra \nu_{f'}|\nu_f(x,t)\ket |^2 \nonumber \\ \nonumber \\
                &=&\sum_{i,j}U_{fi}U_{f'i}U_{fj}U_{f'j}e^{i[(p_i-p_j)x-(E_i-E_j)t]}~. 
\end{eqnarray}
$P_{f \to f'}(x,t)$ with $f'=f$ $(f'\neq f)$ is often called the survival (transition) probability. 

  In the same-energy prescription, one assumes $E_1 =E_2 = \dots =E_N \equiv \tilde{E} $, which leads to
%%%(5)
\begin{equation}
P_{f\to f'}(x,t)=\sum_iU_{fi}^2U_{f'i}^2+2\sum_{i<j}U_{fi}U_{f'i}U_{fj}U_{f'j}\cos (2\pi x/\ell_{ij}) ~,
\end{equation}
where
%%%(6)
\begin{equation}
\left. 
	\begin{array}{ll}
	\ell _{ij}/2\pi =1/|\tilde{p_i}-\tilde{p_j}|=(\tilde{p_i}+\tilde{p_j})/|m_i^2-m_j^2|~, \\ ~\\ 
	\tilde{p_i}=\sqrt{\tilde{E}^2-m_i^2}~,
	\end{array}
\right. 
\end{equation}
while in the same-momentum prescription, one assumes $p_1 = p_2 = \dots = p_N \equiv \tilde{p} $, which leads to
%%%(7)
\begin{equation}
P_{f\to f'}(x,t)=\sum_iU_{fi}^2U_{f'i}^2+2\sum_{i<j}U_{fi}U_{f'i}U_{fj}U_{f'j}\cos (2\pi t/\tau_{ij}) ~,
\end{equation}
where
%%%(8)
\begin{equation}
\left. 
	\begin{array}{ll}
	\tau _{ij}/2\pi =1/|\tilde{E_i}-\tilde{E_j}|=(\tilde{E_i}+\tilde{E_j})/|m_i^2-m_j^2|~, \\ ~\\ 
	\tilde{E_i}=\sqrt{\tilde{p}^2+m_i^2}~.
	\end{array}
\right. 
\end{equation}

  For relativistic neutrinos, approximating both $\tilde{p_i}$ and $\tilde{p_j}$ by $\tilde{E}$ in Eq.(6) and both $\tilde{E_i}$ and $\tilde{E_j}$ by $\tilde{p}$ in Eq.(8), one has
%%%(9)(10)(11)
\begin{eqnarray}
&&\ell_{ij}/2\pi \simeq 2\tilde{E}/|m_i^2-m_j^2|~, \\ 
&&\tau_{ij}/2\pi \simeq 2\tilde{p}/|m_i^2-m_j^2|~, \\ 
&&\tau_{ij}\simeq \ell_{ij} ~,
\end{eqnarray}
and, by arguing that $t$ may be identified with $x$ in Eq.(7) (or $x$ with $t$ in Eq.(5)), it is often claimed that the two prescriptions give practically the same results.

  It is to be emphasized however that, conceptually, the two prescriptions are distinct and Eq.(7) has to be distinguished from Eq.(5). The reason why we insist this will become clear in later sections. To distinguish $\tau_{ij}$ from $\ell_{ij}$, we shall refer to the former as oscillation period (and to the latter as oscillation length).
  
  Importance to distinguish between the two prescriptions was also noted by Lipkin.$^6$ The question as to which is a more appropriate prescription is raised and arguments in favor of the same-energy prescription are given. Furthermore, it is claimed that the same-momentum prescription describes "non-experiments" and it is pointed out that, if difference in arrival time (i.e. the time for the the neutrinos to travel from the creation point to the detection point) were taken into account in Eq.(7), a paradox would arise.$^7$
%%%%%% Sect.3. %%%%%%%%%%%%%%%%

\section{Wave-packet treatments  --- Conditions for oscillation to occur ---}

  In the conventional formalism reviewed above, each mass eigenstate is represented by a plane wave and the formalism is quite far from particle picture. We now, following Kayser,$^5$ introduce $a_i(p)$, which may be interpreted as the amplitude for creation of $\nu_i$ with momentum $p$,$^8$ and replace Eq.(3) by
%%%(12)
\begin{equation}
|\nu_f(x,t)\ket =\sum_iU_{fi}b_i(x,t)|\nu_i\ket ~,
\end{equation}
where
%%%(13)
\begin{equation}
b_i(x,t)=\int_{-\infty}^{\infty}dp\, a_i(p)e^{i(px-E_it)}~.
\end{equation}
To have a wave localized in space with width $\Delta\bar{x}>0$, we take, for simplicity and definiteness,$^9$
%%%(14)
\begin{equation}
a_i(p) = 
\left\{ 
	\begin{array}{ll}
	1/\Delta \bar{p} & ,~|p-\bar{p}| \le \Delta \bar{p}/2 \\ 
	0 & ,~|p-\bar{p}| > \Delta \bar{p}/2 
	\end{array}
\right. 
~,
\end{equation}
where
%%%(15)
\begin{equation}
\Delta\bar{p} \simeq 1/\Delta \bar{x} ~,
\end{equation}
\[
\bar{p}\ge \Delta \bar{p}/2 ~.
\]

Expanding $E_i=E_i(p)=\sqrt{p^2+m_i^2}$ around $\tilde{p_i}$ satisfying   $|\tilde{p_i} - \bar{p}|\le \Delta\bar{p}/2$, and noting
%%%()
\[
px-E_it\simeq \tilde{p_i}x-\tilde{E_i}t+(p-\tilde{p_i})(x-\tilde{\beta_i}t) ~, 
\]
after some algebra, one arrives at
%%%(16)
\begin{equation}
b_i(x,t) = e^{i(\tilde{p_i}x-\tilde{E_i}t)}g_i(x,t) ~,
\end{equation}
where 
%%%(17)
\begin{eqnarray}
g_i(x,t)&=&e^{i(\bar{p}-\tilde{p}_i)(x-\tilde{\beta}_it)}\frac{\sin [(\Delta \bar{p})(x-\tilde{\beta_i}t)/2]}{(\Delta \bar{p})(x-\tilde{\beta_i}t)/2} \nonumber \\ 
&\simeq& \frac{\sin [(\Delta \bar{p})(x-\tilde{\beta_i}t)/2]}{(\Delta \bar{p})(x-\tilde{\beta_i}t)/2}~,
\end{eqnarray}
\[
\tilde{E_i}=\sqrt{\tilde{p_i}^2+m_i^2}~,\qquad \tilde{\beta_i}=\tilde{p_i}/\tilde{E_i} ~. 
\]
  The necessary conditions for $\nu_i$ and $\nu_j$ (with $i < j$) to interfere with each other in the same-energy prescription are
%%%(18)
\begin{equation}
E_i(\hbox{max}) > E_j(\hbox{min}) ~,
\end{equation}
%%%(19)
\begin{equation}
(\Delta\bar{p})|x-\tilde{\beta}_it|  \simeq (\Delta\bar{p})|x-\tilde{\beta}_jt|  \lesssim 1 ~,
\end{equation}
where
\[
E_i(\hbox{max},\hbox{min})=\sqrt{(\bar{p} \pm \Delta \bar{p}/2)^2+m_i^2}~.
\]
Only if Eq.(18) is satisfied, given $\tilde{E}$ satisfying $E_i(\hbox{max}) > \tilde{E} > E_j(\hbox{min})$, one can have $\tilde{p}_i$  and $\tilde{p}_j$ satisfying $\sqrt{\tilde{p}_i^2+m_i^2}=\sqrt{\tilde{p}_j^2+m_j^2}=\tilde{E}$, while Eq.(19) ensures the two mass eigenstates to contribute with appreciable and comparable weight. Eq.(18) gives 
%%%(20)
\begin{equation}
\Delta\bar{p} > |m_i^2-m_j^2|/2\bar{p}~,
\end{equation}
and Eq.(19) gives
%%%(21)
\begin{equation}
(\Delta \bar{p})|\tilde{\beta}_i t-\tilde{\beta}_j t|\lesssim 1~. 
\end{equation}
These two equations together lead to 
%%%(22)
\begin{equation}
|\tilde{\beta}_i t-\tilde{\beta}_j t|~ \lesssim \Delta \bar{x} \lesssim \bar{\ell}_{ij}/2\pi~,
\end{equation}
where
%%%(23)
\begin{equation}
\bar{\ell}_{ij}/2\pi = 2\bar{p}/|m_i^2-m_j^2|~.
\end{equation}
$\bar{\ell}_{ij}$ may be regarded as some mean value of the oscillation length $\ell _{ij}$ defined by Eq.(6).

To develop wave packet treatment appropriate to the same-momentum prescription, one has to introduce $a_i(E)$, the amplitude to create $\nu_i$ with energy $E$. With $a_i(E)$ taken as
%%%(24)
\begin{equation}
a_i(E) = 
\left\{ 
	\begin{array}{ll}
	1/\Delta \bar{E} &,~ |E-\bar{E}| \le \Delta \bar{E}/2 \\ 
	0                &,~ |E-\bar{E}|  >  \Delta \bar{E}/2  
	\end{array}
\right. 
~,
\end{equation}
which implies from uncertainty relation that the waves are localized in time and has width
%%%(25)
\begin{equation}
\Delta \bar{t} \simeq 1/ \Delta \bar{E} ~,
\end{equation}
$b_i(x,t)$ in Eq.(12) is now given by
%%%(26)
\begin{eqnarray}
b_i(x,t)&=&\frac{1}{\Delta\bar{E}}\int_{\bar{E}-\Delta \bar{E}/2}^{\bar{E}+\Delta \bar{E}/2}dE \, e^{i(p_ix-Et)},
\end{eqnarray}
where
\[
p_i = \sqrt{E^2 - m_i^2}~.
\]
Proceeding almost the same way as that leading to Eqs.(16) and (17), one arrives again at Eq.(16), but with $g_i(x,t)$ given this time by
%%%(27)
\begin{eqnarray}
g_i(x,t)& = &e^{-i(\bar{E}-\tilde{E_i})(t-x/\tilde{\beta_i})}\frac{\sin [(\Delta \bar{E})(t-x/\tilde{\beta_i})/2]}{(\Delta \bar{E})(t-x/\tilde{\beta_i})/2} \nonumber\\ 
&\simeq & \frac{\sin [(\Delta \bar{E})(t-x/\tilde{\beta_i})/2]}{(\Delta \bar{E})(t-x/\tilde{\beta_i})/2} ~.
\end{eqnarray}
The necessary conditions for $\nu_i$ and $\nu_j$ (with $i < j$) to be involved in oscillation in the same-momentum prescription are
%%%(28)
\begin{equation}
p_i(\hbox{min}) < p_j(\hbox{max})~,
\end{equation}
%%%(29)
\begin{equation}
(\Delta \bar{E})|t-x/\tilde{\beta}_i| \simeq (\Delta \bar{E})|t-x/\tilde{\beta}_j| \lesssim 1 ~,
\end{equation}
where
\[
p_i(\hbox{max},\hbox{min}) = \sqrt{(\bar{E}\pm \Delta \bar{E}/2)^2 - m_i^2}~.
\]
These two equations together lead to 
%%%(30)
\begin{equation}
|x/\tilde{\beta}_i-x/\tilde{\beta}_j| \lesssim \Delta \bar{t} \lesssim \bar{\tau}_{ij}/2\pi~,
\end{equation}
where
%%%(31)
\begin{equation}
\bar{\tau}_{ij}/2\pi = 2\bar{E}/|m_i^2-m_j^2|~.
\end{equation}
$\bar{\tau}_{ij}$ may be regarded as some mean value of the oscillation period $\tau _{ij}$ defined by Eq.(8).

%%%%%%%% Sect.4. %%%%%%%%%%%%%5

\section{Some phenomenological implications}

  The theoretical expressions for $P_{f \to f'}(x,t)$ now read: 
%%%(32)
\begin{equation}
P_{f\to f'}(x,t)=\sum_i U_{fi}^2U_{f'i}^2 f_{ii}(x,t)+2\sum_{i<j}U_{fi}U_{f'i}U_{fj}U_{f'j}f_{ij}(x,t) \cos (2\pi x/ \ell _{ij})~,
\end{equation}
or
%%%(33)
\begin{equation}
P_{f\to f'}(x,t)=\sum_i U_{fi}^2U_{f'i}^2 f_{ii}(x,t)+2\sum_{i<j}U_{fi}U_{f'i}U_{fj}U_{f'j}f_{ij}(x,t) \cos (2\pi t/ \tau _{ij})~,
\end{equation}
where
%%%(34)
\begin{equation}
f_{ij}(x,t)=g_i(x,t)g_j(x,t),
\end{equation}
with $g_i(x,t)$ given either by Eq.(17) or by Eq.(27). 

  The presence of the factor $f_{ij}(x,t)$, which depends in general on the suffices $i$ and $j$, should be taken into account in phenomenological analyses, e.g. extraction of the mixing matrix elements from observed oscillation pattern, interpretation in terms of neutrino oscillation of observed deficit or depletion of some kind of neutrino flux as compared to some expectation, and so on. It would also cause, in particular in the case that non-relativistic neutrinos are involved, this or that mass components to drop out of interference.$^{11}$

  For relativistic neutrinos, the difficulties which could arise from presence of the factor $f_{ij}(x,t)$ may be largely reduced. The same-energy and same-momentum prescriptions will practically have no difference, as already mentioned and have been well realized, and the first inequality in Eq.(22) or in Eq.(30) will be trivially satisfied, leaving the second inequality in Eq.(22) or in Eq.(30) as the only condition for oscillation to occur.

%%%%%% Sect.5. %%%%%%%%%%%%

\section{The same-velocity prescription}

Instead to suppose $E_1 = E_2=\dots =E_N$ or $p_1 = p_2=\dots =p_N$, let us suppose
\[
\beta_1=\beta_2=\dots =\beta_N \equiv \tilde{\beta}~ .
\] 
We have then from Eq.(4)
%%%(35)
\begin{equation}
P_{f\to f'}(x,t)=\sum_i U_{fi}^2U_{f'i}^2+2\sum_{i<j}U_{fi}U_{f'i}U_{fj}U_{f'j}\cos (2\pi (x-t/\tilde{\beta})/\ell '_{ij})~,
\end{equation}
where
%%%(36)
\begin{equation}
\ell'_{ij}/2\pi =1/|\tilde{p}_i-\tilde{p}_j|=1/\tilde{\beta}\tilde{\gamma}|m_i-m_j|= (\tilde{p}_i + \tilde{p}_j)/\tilde{\beta}^2\tilde{\gamma}^2|m_i^2 - m_j^2|~,
\end{equation}

\[
\tilde{p}_i=\tilde{\beta}\tilde{E}_i=\tilde{\beta}\tilde{\gamma}m_i,~~~~\tilde{\gamma}=1/\sqrt{1-\tilde{\beta}^2}~.
\]

 The wave packet treatment as that leading to Eq.(32) modifies Eq.(35) into
%%%(37)
\begin{equation}
P_{f\to f'}(x,t)=\sum_{i}U_{fi}^2U_{f'i}^2 f_{ii}(x,t) +2\sum_{i<j}U_{fi}U_{f'i}U_{fj}U_{f'j}f_{ij}(x,t)\cos (2\pi (x-t/\tilde{\beta})/\ell '_{ij})~,
\end{equation}
where $f_{ij}(x,t)$ is again defined by Eq.(34) with $g_{i}(x,t)$ given by
%%%(38)
\begin{eqnarray}
g_i(x,t)&=&e^{i(\bar{p}-\tilde{p}_i)(x-\tilde{\beta}t)}\frac{\sin [(\Delta \bar{p})(x-\tilde{\beta}t)/2]}{(\Delta \bar{p})(x-\tilde{\beta}t)/2} \nonumber \\ 
&\simeq& \frac{\sin [(\Delta \bar{p})(x-\tilde{\beta}t)/2]}{(\Delta \bar{p})(x-\tilde{\beta}t)/2}~.
\end{eqnarray}
One may convince himself that the necessary conditions for $\nu_i$ and $\nu_j$ (with $i < j$) to interfere in this third prescription are
%%%(39,40)
\begin{equation}
\beta_i(\hbox{min}) < \beta_j(\hbox{max}) ~,
\end{equation}
\begin{equation}
(\Delta \bar{p})|x-\tilde{\beta}t| \lesssim 1 ~,
\end{equation}
where
\[
\beta_i(\hbox{max},\hbox{min}) = (\bar{p}\pm\Delta\bar{p}/2)/\sqrt{(\bar{p}\pm\Delta\bar{p}/2)^2 + m_i^2}~.
\]
Eq.(39) gives
%%%(41)
\begin{equation}
\Delta \bar{x} \lesssim \bar{\ell} '_{ij}/2\pi~,
\end{equation}
where
%%%(42)
\begin{equation}
\bar{\ell} '_{ij}/2\pi=(m_i^2+m_j^2)/|m_i^2-m_j^2|\bar{p}~.
\end{equation}
$\bar{\ell}'_{ij}$ is related to the oscillation length $\ell '_{ij}$ defined by Eq.(36) as
\[
\bar{\ell} '_{ij}=[(\tilde{p}_i^2+\tilde{p}_j^2)/\bar{p}(\tilde{p}_i+\tilde{p}_j)] \ell '_{ij}~,
\] 
and again may be regarded as some mean value of $\ell '_{ij}$.
 
  A couple of comments are in order.
  
(1) The condition corresponding to Eq.(21) is always satisfied exactly;

(2) The factor $f_{ij}(x,t)$ in Eq.(37) is actually independent of the suffices $i$ and $j$;

(3) There is no room for such a paradox as Lipkin$^6$ mentioned to arise;

(4) If advantage is taken of $g_{i}(x,t)$ (and hence $f_{ij}(x,t)$) having a sharp peak at $x - \tilde{\beta}t=0$ in Eq.(37), one has
%%%(43)
\begin{equation}
P_{f\to f'}(x,t)=\sum_{i}U_{fi}^2U_{f'i}^2 f_{ii}(x,t) +2\sum_{i<j}U_{fi}U_{f'i}U_{fj}U_{f'j}f_{ij}(x,t)\cos (2\pi x/\ell ''_{ij})~,
\end{equation}
where
%%%(44)
\begin{equation}
\ell ''_{ij}/2\pi = \tilde{\beta}^2\tilde{\gamma}^2/|\tilde{p}_i - \tilde{p}_j| = \tilde{\beta}\tilde{\gamma}/|m_i - m_j| = (\tilde{p}_i + \tilde{p}_j)/|m_i^2 - m_j^2|~.
\end{equation}
It is amusing to see that $\ell ''_{ij}$ in its last form appears similar to $\ell_{ij}$ defined by Eq.(6).
%%%%%% Sect.6. 

\section{Concluding remarks}

The meaning of the conditions (22) we have derived is quite obvious and it seems that no word needs to be added. 

That, in order for oscillation to take place, some conditions like (22) have to be satisfied has already been noted by Kayser$^5$ and by Lipkin.$^6$ They first point out that a "missing-mass measurement" would prevent oscillation or interference to occur and then develop arguments based on uncertainty relation between $x$ and $p$ to find them. We have put them in a more quantitative and well-defined form, and derived them in a way tied explicitly and closely to wave packet treatments and to relativistic kinematics, though physics behind is of course essentially the same.

The meaning of the conditions (30) is also quite obvious. Physics behind is uncertainty relation between $t$ and $E$, and we have derived them by introducing $a_i(E)$, the amplitude to create $\nu_i$ with energy $E$, and considering wave packet localized in time. In proceeding in this way, we have had quantum mechanics formulated with the time as an observable in mind.$^{14}$ Although such attempts have encountered some difficulties,$^{15}$ our arguments and treatments as a whole clearly indicate that the same-momentum prescription has to be distinguished conceptually from the same-energy prescription. Whether and how experiments to be described by the former prescription could become realistic and feasible remain to be carefully examined and contrived with efforts. This last comment applies also to experiments to be described by the same-velocity prescription.

In conclusion, we like to recall again that involved in neutrino oscillation are such  important physics as superposition principle and uncertainty relation of quantum mechanics, and relativistic kinematics as well, and express our hope that our discussions presented here, though with emphases placed more on conceptual aspects, would have some relevance to current and future experimental as well as phenomenological studies on neutrino oscillation.

%%%% Acknowledgements %%%%%%

\section*{Acknowledgements}

We are grateful to Professor S.~Kamefuchi for an illuminating discussion and to the members of the Particle Physics Group at Nihon University for continuous encouragement.

%%%%% References %%%%%%%


\begin{thebibliography}{100}
\bibitem{7}Lipkin(Ref.~6) showed this for the case of neutral meson (i.e. $K^0-\overline{K^0}$ and $B^0-\overline{B^0}$) oscillations. Note that, if difference in travel distance (distance between the creation point and the detection point of the neutrinos) were taken into account in Eq.(5), a similar paradox would arise.
\bibitem{8}Kayser(Ref.~5) considered the case that the neutrinos  were created as result of pion decays and introduced  $a( p_{\pi}^i(p))$, the amplitude for a parent pion to have the momentum $p_{\pi}^i(p)$ which leads to $\nu_i$ with momentum $p$.
\bibitem{9}The essential points which follow should not depend much on our choice made here. The more realistic case,
\[
a_i(p) = 
\left\{ 
	\begin{array}{ll}
	1/\Delta \bar{p}_i & ,~|p-\bar{p}_i| \le \Delta \bar{p}_i/2 \\ 
	0 & ,~|p-\bar{p}_i| > \Delta \bar{p}_i/2 
	\end{array}
\right. 
~,
\]
and the cases in which other possible functional forms such as Gaussian form are chosen will be discussed separately.(Ref.~10)
\bibitem{11} This possibility has been well recognized implicitly and discussed explicitly for example by Nussinov,(Ref.~12) by Kaser(Ref.~5) and recently by Ahluwalia.(Ref.~13)
\end{thebibliography}
\end{document}